\documentclass{sig-alternate-05-2015}
\usepackage{url}

\usepackage{amsmath}
\usepackage{amssymb}
\usepackage{graphicx}
\usepackage{array}
\usepackage{url}
\usepackage{color}
\usepackage{multirow}
\usepackage{booktabs}
\usepackage{verbatim}
\usepackage[english]{babel}
\usepackage{colortbl}
\usepackage[utf8]{inputenc}

\usepackage{blindtext}
\usepackage{etoolbox}
\makeatletter
\patchcmd{\maketitle}{\@copyrightspace}{}{}{}
\makeatother
\begin{document}



\clubpenalty=10000
\widowpenalty=10000

\title{The Emergence of Crowdsourcing among Pok{\'e}mon Go Players\titlenote{This is a preprint of an article submitted to a conference.}}
%
%
%
%
%

\numberofauthors{4} 
%
\author{
%
%
\alignauthor
Priscila Martins\\
       \affaddr{UFMG, Brazil}\\
       \email{\large primsouza@dcc.ufmg.br}
\alignauthor
Manoel Miranda\\
       \affaddr{UFMG, Brazil}\\
       \email{\large manoelrmj@dcc.ufmg.br}
\and  
\alignauthor
Fabr\'icio Benevenuto\\
       \affaddr{UFMG, Brazil}\\
       \email{\large fabricio@dcc.ufmg.br}
\alignauthor
Jussara Almeida\\
       \affaddr{UFMG, Brazil}\\
       \email{\large jussara@dcc.ufmg.br}
}

\date{April 2017}

\maketitle

\begin{abstract}
Since its launching, Pok{\'e}mon Go has been pointed as the largest gaming phenomenon of the smartphone age. As the game requires the user to walk in the real world to see and capture  Pok{\'e}mons, a new wave of crowdsourcing apps have emerged to allow users to collaborate with each other, sharing where and when Pok{\'e}mons were found. In this paper we characterize one of such initiatives, called PokeCrew.  Our analyses uncover a set of aspects of user behavior and system usage in such emerging crowdsourcing task, helping unveil some problems and benefits. We hope our effort can inspire the design of new crowdsourcing systems.
\end{abstract}

\section{Introduction}  

The mobile games industry experienced an exponential growth in the past decade, motivated mainly by (i) an ever increasing worldwide penetration of smartphones and mobile devices, (ii) the ability of such devices to deliver quality audio and video; and (iii) the increasing capacity  of network transmissions of these devices, allowing users to download larger and more complex games~\cite{Soh:2008:MG:1325555.1325563}. 

The largest gaming phenomenon of the smartphone age so far has been the augmented reality game {\it Pok{\'e}mon Go}. It was launched in July 2016, firstly in Australia, New Zeland, and USA. Yet, in one week after launching, it had already reached seven million users, accounting for three to six times more downloads of the most popular games in history at that time~\cite{pokemongostatistics}. The game makes use of GPS, camera, and position sensors of smartphones which allow its users to capture, battle and train virtual creatures called Pok{\'e}mons. These creatures appear on the phone screen as if they were in the real world. The set of technologies that allow this kind of experience support the so-called augmented reality, a field that has received a lot of attention after the game success.

There has been a number of recent studies exploiting behavioral changes among Pok{\'e}mon Go players. Nigg et al.~\cite{nigg2017Pokemon} suggest that Pok{\'e}mon Go may increase physical activity and decrease sedentary behaviors. Other efforts~\cite{tateno2016new,dorward2016Pokemon, de2016field} argue that the game may represent a new shift in perspective:  players tend to socialize more while playing as they tend to concentrate in popular areas of the game, often  called Pok{\'e}Stops. 

Since the game requires the user to walk in the real world to see and capture the Pok{\'e}mons nearby, a new wave of supporting apps has emerged. 
In these apps, players can collaborate with each other, sharing where and when Pok{\'e}mons were found. They represent the emergence of a crowdsourcing effort of the game players to find rare and valuable Pok{\'e}mons.  PokeCrew\footnote{https://www.pokecrew.com/}, one such app of great popularity, is  a crowdsourced Pok{\'e}mon Go map. It shows reports of locations of Pok{\'e}mon posted by players in real time in a map and it became quite popular among the most active users. For example, this website was ranked among the top 15000 domains in the Web, according to Alexa.com~\cite{alexa} and its IOS and android versions had hundreds of thousands downloads. 

In this paper, we characterize the crowdsourcing effort of Pok{\'e}mon Players through PokeCrew. Crowdsourcing systems enlist a multitude of humans to help solve a wide variety of problems.
Over the past decade, numerous such systems have appeared on the  Web. Prime examples include Wikipedia, Yahoo! Answers, Mechanical Turk-based systems, and  many more~\cite{howe2006rise}. Our effort  consists in characterizing an emerging type of crowdsourcing system, identifying many interesting technical and social challenges. To that end, we obtained a near two-month log of reports from the game players, containing 39,895,181 reports of Pok{\'e}mon locations. Our analyses uncover a set of aspects of user behavior and system usage in an emerging crowdsourcing task. We hope our effort can inspire the design of emerging crowdsourcing systems. 

In the following, we first describe the data used in our study and then analyze how users collaboratively help each other within the Pokecrew platform. We finish this paper with our conclusions and possible directions for future work.

\section{Dataset}

With the increasing popularity of  Pok{\'e}mon Go in the whole world, many applications emerged with the purpose of enhancing the players experience with the game. Among the most popular ones are Pokecrew\footnote{www.pokecrew.com}, PokeRadar\footnote{https://www.Pokemonradargo.com/}, and PokeVision\footnote{www.pokevision.com}. The idea is to share Pok{\'e}mon maps and their locations in a crowdsourced way: after finding Pok{\'e}mons in the game itself users may  report them in the supporting app, making the creatures visible to other players that are not in that specific location and time. This can be very useful to players, since, in the game, it is not possible to see Pok{\'e}mons far from where the player is currently physically located.

We have obtained data from Pokecrew, a popular app that offers to users a map containing the location of Pok{\'e}mons reported by other users. The application can be found in the PlayStore, AppStore, as well as on the Web. 
Our dataset contains 39,895,181 reports, from July 12$^{th}$ to August 24$^{th}$ 2016. Each report contains several information fields, including: a report id, reported Pok{\'e}mon id, geographic coordinates, time when the report was created and, in some registers, a username. 

Our results show that most reports in our dataset (98.7\%) do not include  a valid username, since the app does not require the user to identify itself in order to create reports. Thus, although we report general statistics computed over the whole dataset in the next section, we focus on the subset of reports with valid usernames to study user behavior. We note that, despite the small percentage, there is still a considerable amount of identifiable reports (over 500k)  on which we can perform such analysis. 
Finally, we also note that the spatial information in our dataset refers to  geographic coordinates of the reports. 


\section{Reported Pok{\'e}mons and their Locations}

We start our characterization by analyzing which Pok{\'e}mons are the most reported ones and where they were reported. We then discuss the application adoption on specific locations. 

\subsection{Most Reported Pok{\'e}mons}

Table~\ref{tab:toppok} shows the top-10 most reported Pok{\'e}mons in our dataset. We  note that these Pok{\'e}mons  are evolutions or difficult  to find in the game. An evolution of a given Pok{\'e}mon consists of a similar monster but with a higher power, which is very important to battle with other Pok{\'e}mons in the so called 'gyms', popular places across the world where a user (a Pok{\'e}mon master) battles with other users in order to take control of that gym. Besides that, having these evolutions contributes to the user's Pokedex, which is a list containing detailed stats for every creature from the Pok{\'e}mon games. The more Pok{\'e}mons a user has in her list, more experience she has on the game, which is also important to battle with other players at the gyms. Capturing an evolution is attractive to  players because the only alternative way to obtain them is to use an egg, which is earned as the player progress in the game. With this egg, the Pok{\'e}mon master can put it to crash, which is achieved by walking. The distance required to crash an egg may vary (2, 5 or 10 kilometers). The higher the distance, the more valuable the Pok{\'e}mon that comes out of the egg is. Thus, it is much more convenient to catch the evolved Pok{\'e}mon right away than it is to walk waiting for the egg to crack and, luckily, be rewarded with a powerful Pok{\'e}mon. 

This observation shows the great contribution of Pokecrew to Pok{\'e}mon players:  the interest in rare Pok{\'e}mons or evolutions is what drive players to appeal to these crowdsourced apps. These Pok{\'e}mons are more valuable than the most commonly found, which normally have less power to battle.  

    \begin{table}[ht]
      \centering
      \begin{tabular}
        {l|r}
        \hline	
        Pok{\'e}mon's Name      & Number of Reports \\
        \hline
        Fearow              & 4,957,913 \\
        Raichu              & 3,910,515 \\
        Slowbro             & 2,395,383 \\
        Golduck             & 2,150,770 \\
        Pidgeot             & 2,051,487 \\
        Nidoran             & 1,507,741 \\
        Tentacool           & 1,199,451 \\
        Nidoqueen           & 1,044,395 \\
        Magnemite           & 938,829 \\
        Clefable            & 928,105 \\
        \hline
      \end{tabular}
      \caption{Top 10 reported Pok{\'e}mons}
      \label{tab:toppok}
      \label{reported Pokemons}
    \end{table}

\subsection{Pok{\'e}mon Location}

We note that our dataset contains only geographic coordinates of the reports. In order to characterize the location where these reports were made,  we first converted the coordinates to the cities and countries where the reports are made.
Our approach to do that consisted of using a reliable Python library, namely geopy\footnote{https://pypi.python.org/pypi/geopy/}, which allows us to retrieve the nearest town/city for a given latitude/longitude coordinate.  

Most of the reports (almost 34\%) come from the U.S, followed by Singapore and Malyasia.  We also noticed that three American cities -- New York City, San Francisco and Santa Monica -- are included in the top 10 cities.


\subsection{Crowdsourcing Adoption}

In this kind of application, the initial stage of its lifespan can be unattractive to the users due to the lack of data in the system. In PokeCrew and other competing apps, this aspect is even more critical, since the adoption of Pok{\'e}mon GO  was fast and  user engagement was very strong. In an application like Pokecrew, if the user does not encounter Pok{\'e}mon reports, it is very likely she will not be motivated to use the system and therefore won't be encouraged to create new reports.

\begin{figure}[!tbh]
\centering
\includegraphics[scale=0.45]{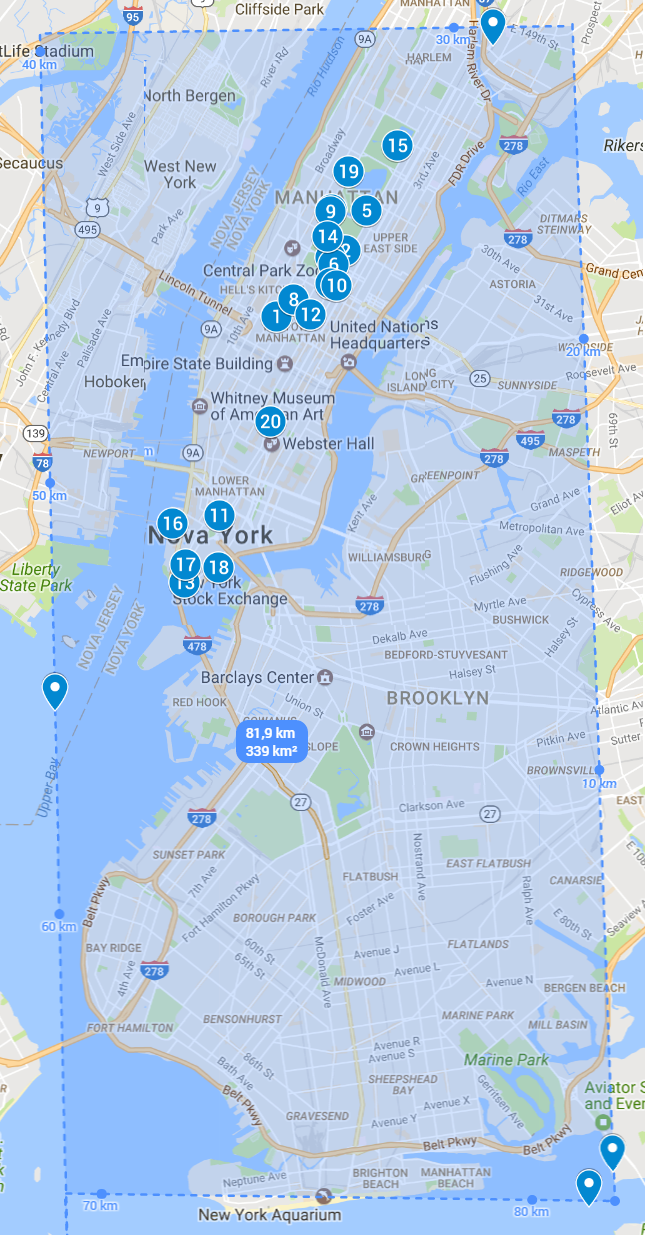}
\caption{Geographical Location of Top 20 regions from NY}
\label{fig:NY20regions}
\end{figure}

\begin{figure}[!thb]
\centering
\includegraphics[scale=0.48]{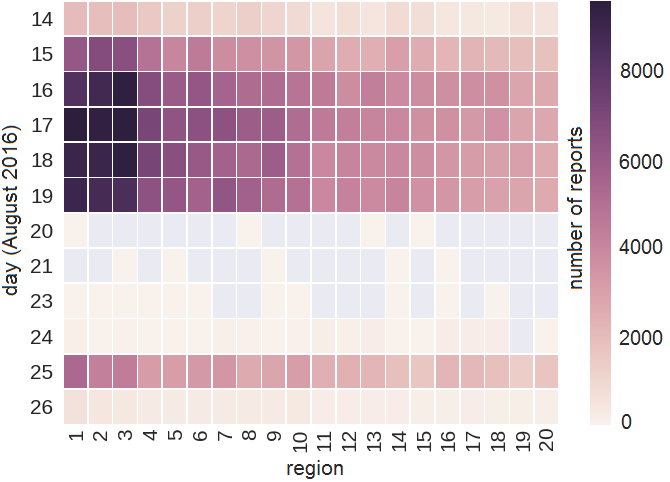}
\caption{Heat Map of Top 20 regions from NY}
\label{fig:heatmap}
\end{figure}

To assess whether the previous presence of Pok{\'e}mons in the application influences the user to create new reports, we compared the amount of reports in a popular area among the days. To that end, we focused on the reports in the city of New York, which concentrates most part of reports. The city geographic area was first divided into 800 regions of approximately $500m^2$ , and for each region we counted the total number of reports created on each day. Besides that, we considered the period from August 14$^{th}$ to 26$^{th}$, which concentrates a larger number of reports. It is possible to see in Figure \ref{fig:NY20regions}, most of the reports made in New York City, were created in the Central Park area, a very popular place in the game itself. Figure~\ref{fig:heatmap} shows a heat map correlating the number of reports in each region showed in Figure~\ref{fig:NY20regions} in each date.

\section{User Behavior}

Next, we provide a characterization of the crowdsourced Pokecrew data, exploring aspects of user behavior and their engagement within this crowdsourced system. The dataset collected presents 39,895,181 reports, but only  452,359 (1.3\%) are registered users (non-anonymous). We focus the next analysis on the behavior of this identified group of users.

\subsection{User Engagement}

Analyzing the reports made by registered users, we can notice that some of them, mostly users who contributed with the largest amounts of  reports, reported a large number of sightings in just one day. For example, the top 1 user reported 184,615 times, being her reports concentrated between July 21$^{st}$ and 24$^{th}$. Specifically, reported 13,007 times on July 21$^{st}$ and 75,574 on July 22$^{nd}$, which are very expressive numbers.
Similarly, 31,656 reports made by the second most active user, which corresponds to almost 99.7\% of his contribution to the Pokecrew system,  were concentrated on a single day, August 12$^{th}$. 
The 10 most active users in the Pokecrew system, in terms of reports of Pok{\'e}mon sightings, are shown in Table \ref{tab:top10users}. We note that some users have reported far many sightings than others, especially the 4 most active ones. Given the large amount of reports associated with these users, and the short time interval during which they were made, we speculate that  these reported sightings may not have been made by ``legitimate" Pokecrew users.


    \begin{table}[!b]
      \centering
      \begin{tabular}
        {p{0.4\linewidth}p{0.4\linewidth}}
        \hline	
        Counting reports    & User name \\
        \hline
        184615              & yay1199 \\
        31766               & hi \\
        609                 & mongrelo \\
        413                 & luchocadaingles \\
        270                 & maestroPok{\'e}monbelloto \\
        261                 & rooty \\
        255                 & srdandrea \\
        239                 & crescenttough \\
        236                 & ceryatec \\
        235                 & guantoresp \\
        \hline
      \end{tabular}
      \caption{Top 10 users and their counting reports}
      \label{tab:top10users}
    \end{table}
    
Based on the amount of reports that the top 5 have made, and the fact that these reports are, in general, concentrated in a few set of days, we've disregarded this data for some analysis. Just a few number of users have reported from 30 to 261 times. Most of users have reported from 1 to 20 times. We have categorized users who have reported from 1 to 5 times as less active users, representing 80\% of the database.

Given that for each report we have the information about the time when it was created and its location, we can calculate for each pair of reports of a single user the speed in which he would have to dislocate in order to make both reports. Calculating the speed for each pair of user's report, we obtained a set of speeds from the identified user's reports. The chart bellow shows the speed distribution across the total identifiable users:

\begin{figure}[!htb]
\centering
\includegraphics[scale=0.43]{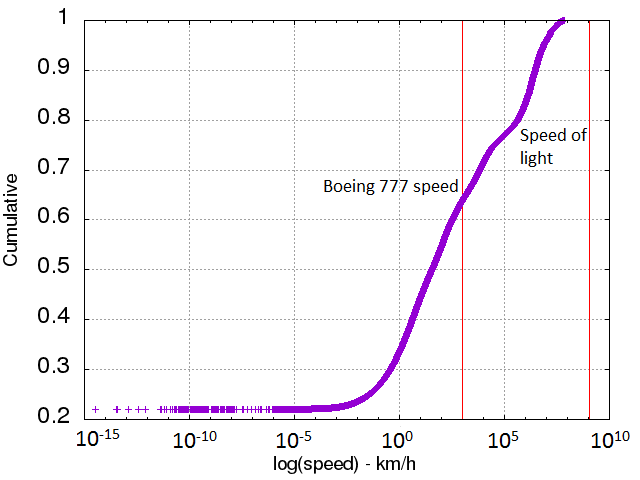}
\caption{Speed distribution across identified users}
\end{figure}

As we can see in the plot above, a considerable number of reports made by the users in the system reveal that they would have to move at abnormally high speeds for this kind of game. Even if we consider the case where a user report a Pok{\'e}mon in one place, go on an airplane trip and then report another in the destination, which although maybe uncommon, is possible, some speeds are not feasible to achieve. The problem is that, even though the PokeCrew app uses the GPS data from the smartphone and places the map in the user current location, the player has the ability to move the map to any place in the world and therefore report a sighting from anywhere. There is no validation in the app if the report being created is trustful. This opens a serious flaw in the system, because malicious users and even bots could create fake sightings to spoof the system and degrade the legit user experience. Even if the user could not change the location in the map, it would still be possible to report a Pok{\'e}mon that does not exist in the game itself at that given time and location. This kind of validation is a key problem in collaborative systems, and can be very challenging due to the lack of mechanisms to control whether the information being supplied to the database is true or not.

\subsection{Temporal Analysis}

The created at field represents the date when the the sighting was reported by a user. The chart below shows an analyse about how many reports were made in each day. From July 12$^{th}$, when Pokecrew was created, to August 7$^{th}$ there were not many users contributing to the system. 
This scenario changes after August 8$^{th}$, with a peak of reports on August 21$^{st}$.

\begin{figure}[!htb]
    \centering
    \includegraphics[width=8.2cm]{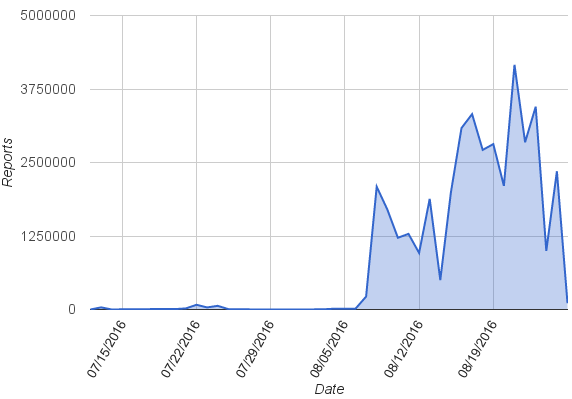}
    \caption{Amount of reports per date}
\end{figure}

\section{Concluding Discussion}

This paper explores one out of a new wave of applications that rely on collaborative databases in the mobile gaming environment. Pok{\'e}mon Go was a huge phenomenon in this area and motivated the creation of innumerable initiatives to help gamers with the task of catching, training and battling Pok{\'e}mons across the city. Since the game requires the user to walk and visit places in order to succeed at being a Pok{\'e}mon master, this kind of support showed to be extremely useful to players, because they could go straight to the exact location where a desired Pok{\'e}mon is located instead of randomly walk hoping to find some valuable monster, as we could see with the most reported Pok{\'e}mons. 
One important aspect in this work was to show how vulnerable such systems are to spoofed data. Inconsistent locations over time for some users and the noticeably high amount of reports these users pushed to the system certainly had a negative impact over the legit final user experience, who could encounter fake reports on the map. Although the app lacks mechanisms to prevent these fake reports, the establishment of a trustful relation with the user is difficult to achieve

\section{Acknowledgments}

We would like to thank Pokecrew for kindly sharing its data with our research group. This work is supported by author's individual grants from Capes, Fapemig, and CNPq. F. Benevenuto is also supported by Humboldt Foundation.



\end{document}